\documentclass[pdftex,twocolumn,epjc3]{svjour3}          

\RequirePackage[T1]{fontenc}

\smartqed  

\RequirePackage{graphicx}
\RequirePackage{mathptmx}      
\RequirePackage{flushend}
\RequirePackage[numbers,sort&compress]{natbib}
\RequirePackage[colorlinks,citecolor=blue,urlcolor=blue,linkcolor=blue]{hyperref}

\journalname{Eur. Phys. J. C}

\usepackage{blindtext}
\usepackage{graphicx}
\usepackage{amsmath}
\usepackage{amssymb}
\usepackage{color}
\usepackage[tight]{units}

\begin{document}
\title{Performance of a CRESST-II Detector Module with True $4\pi$-veto}

\author{G. Angloher\thanksref{MPP}
        \and
       P. Bauer\thanksref{MPP}
       \and
       A. Bento\thanksref{MPP,COIM}
       \and
       C. Bucci\thanksref{LNGS}
       \and
       L. Canonica\thanksref{LNGS,MIT}
       \and
       X. Defay\thanksref{TUM}
       \and
       A. Erb\thanksref{TUM,BADW}
       \and
       F. v. Feilitzsch\thanksref{TUM}
       \and
       N. Ferreiro-Iachellini\thanksref{MPP}
       \and
       P. Gorla\thanksref{LNGS}
       \and
       A. G\"{u}tlein\thanksref{VIENNA1,VIENNA2}
       \and
       D. Hauff\thanksref{MPP}
       \and
       J. Jochum\thanksref{TUE}
       \and
       M. Kiefer\thanksref{MPP}
       \and
       H. Kluck\thanksref{VIENNA1,VIENNA2}
       \and
       H. Kraus\thanksref{OX}
       \and
       J.-C. Lanfranchi\thanksref{TUM}
       \and
       A. Langenk\"{a}mper\thanksref{TUM}
       \and
       J. Loebell\thanksref{TUE}
       \and
       M. Mancuso\thanksref{MPP}
       \and
       A. M\"{u}nster\thanksref{TUM}
       \and
       C. Pagliarone\thanksref{LNGS}
       \and
       F. Petricca\thanksref{MPP}
       \and
       W. Potzel\thanksref{TUM}
       \and
       F. Pr\"{o}bst\thanksref{MPP}
       \and
       F. Reindl\thanksref{MPP,ROMA}
       \and
       J. Rothe\thanksref{MPP}
       \and
       K. Sch\"{a}ffner\thanksref{LNGS,GSSI}
       \and
       J. Schieck\thanksref{VIENNA1,VIENNA2}
       \and
       V. Schipperges\thanksref{TUE}
       \and
       S. Sch\"{o}nert\thanksref{TUM}
       \and
       W. Seidel\thanksref{TUM,t1}
       \and
       M. Stahlberg\thanksref{VIENNA1,VIENNA2}
       \and
       L. Stodolsky\thanksref{MPP}
       \and
       C. Strandhagen\thanksref{TUE}
       \and
       R. Strauss\thanksref{MPP}
       \and
       A. Tanzke\thanksref{MPP}
       \and
       H.H. Trinh Thi\thanksref{TUM}
       \and
       C. T\"{u}rkoglu\thanksref{VIENNA1,VIENNA2}
       \and
       M. Uffinger\thanksref{TUE}
       \and
       A. Ulrich\thanksref{TUM}
       \and
       I. Usherov\thanksref{TUE}
       \and
       S. Wawoczny\thanksref{TUM}
       \and
       M. Willers\thanksref{TUM}
       \and
       M. W\"{u}strich\thanksref{MPP,e1}
       \and
       A. Z\"{o}ller\thanksref{TUM}
}

\thankstext[$\dag$]{t1}{deceased (20/02/17)}
\thankstext[$*$]{e1}{Corresponding author: marc.wuestrich@mpp.mpg.de}

\institute{Max Planck Institute f. Physics, 80805 Munich, Germany\label{MPP}
          \and
          Laboratori Nazionali del Gran Sasso, 67100 Assergi L'Aquila, Italy\label{LNGS}
          \and
          Physics Department and Excellence Cluster Universe, Technische Universit\"{a}t M\"{u}nchen, 85747 Garching, Germany\label{TUM}
          \and
          Institut f. Hochenergiephysik der \"{O}sterreichischen Akademie der Wissenschaften, A-1050 Vienna, Austria\label{VIENNA1}
          \and
          Atominstitut, Vienna University of Technology, A-1020 Vienna, Austria\label{VIENNA2}
          \and
          Eberhard-Karls-Universit\"{a}t T\"{u}bingen, 72076 T\"{u}bingen, Germany\label{TUE}
          \and
          Department of Physics, University of Oxford, Oxford OX1 3RH, United Kingdom\label{OX}
          \and
          Also at: Departamento de Fisica, Universidade de Coimbra, P3004 516 Coimbra, Portugal\label{COIM}
          \and
          Also at: Massachusetts Institute of Technology, Cambridge, MA 02139, USA\label{MIT}
          \and
          Also at: Walther-Mei\ss{}ner-Institut f. Tieftemperaturforschung, 85748 Garching, Germany\label{BADW}
          \and
          Also at: GSSI-Gran Sasso Science Institute, 67100, L'Aquila, Italy\label{GSSI}
          \and
          \emph{Present Address:}  INFN-Sezione di Roma, 00185 Rome, Italy\label{ROMA}
}

\maketitle

\begin{abstract}
 Scintillating, cryogenic bolometers are widely used in the field of rare event searches.
 Their main advantages are an excellent energy resolution and particle identification on an event-by-event basis. 
 The sensitivity of experiments applying this detector technique can be limited by the performance of the light channel and the presence of external backgrounds in the region of interest.
 In the framework of the CRESST-II experiment, we developed and successfully tested a novel detector design addressing both challenges.
 Using a large scale ($\approx$\unit[60]{cm$^2$}), beaker-shaped silicon light absorber, the signal height recorded in the light channel is improved by a factor 2.5 compared to conventional CRESTT-II detector modules.
 In combination with a large carrier crystal, a true $4\pi$ veto system is established which allows to tag external background sources.
\end{abstract}

\section{Introduction}

Experiments for direct dark matter searches (e.g. CRESST-II\cite{Lise_Results}, EDEWEISS\cite{Hehn2016}, CDMS-II\cite{CDMS2016}) as well as experiments aiming at the detection of the neutrinoless double beta decay ($0\nu\beta\beta$-decay) (e.g. CUORE\cite{Alduino2017}, CUPID-0\cite{Artusa2016}, AMORE\cite{Amore}) apply cryogenic detector techniques to reach their scientific goals.

In a cryogenic scintillating bolometer (``phonon-light technique''), a scintillating target crystal is equipped with a highly sensitive thermometer\footnote{Typically transition edge sensors (TES) or neutron transmutation doped thermistors (NTD)} to measure the tiny temperature rises induced by the interacting particles (phonon signal) and is paired with a spatially separated light detector which detects emitted scintillation light in coincidence (light signal).
The phonon signal provides precise and particle-independent measurement of the total energy deposition in the target crystal. 
The amount of scintillation produced simultaneously during an energy deposition strongly depends on the type of interacting particle.
Thus, the light signal allows to distinguish nuclear recoils from interactions of $\beta$/$\gamma$ and $\alpha$ particles happening inside the target crystal. 

Two aspects of the detector concept limit the sensitivity of experiments applying it.\\
Although the performance of the light channel usually exceeds the performance of the phonon channel in terms of sensitivity and resolution, the small energy share which is emitted in form of scintillation light represents a limit to the particle identification capabilities for small energy depositions ($<\mathcal{O}(\unit[10]{keV})$).\\
While the identification of events originating from the bulk of the target crystal is achieved unambiguously, events with different origin are able to create unvetoed background contributions (e.g. recoils from $\alpha$-decays on nearby surfaces in case of dark matter searches).
To identify these events complex detector housing concepts are needed which allow the unambiguous identification of all event classes \cite{TUM40_surface_rejection}.

In the framework of the CRESST-II dark matter search a new detector design was developed which overcomes both of the aforementioned limitations of the phonon-light technique.
In this paper we explain in detail the features of the detectors establishing a complete $4\pi$-veto system.
We report and evaluate the performance of this new detector concept acquired from two prototype modules installed in CRESST-II (Phase 2) taking data for two years in the Laboratori Nazionali del Gran Sasso (LNGS). 

\section{Technical Description}

Developed in the CRESST-II framework, the absorber material used for this prototype module is a calcium tungstate crystal (CaWO$_{4}$).
A cylindrically shaped target crystal with a diameter of \unit[35]{mm} and a height of \unit[38]{mm} is used as main absorber. ($\approx\unit[220]{g}$).

\begin{figure}
\centering
\includegraphics[width=1.0\linewidth]{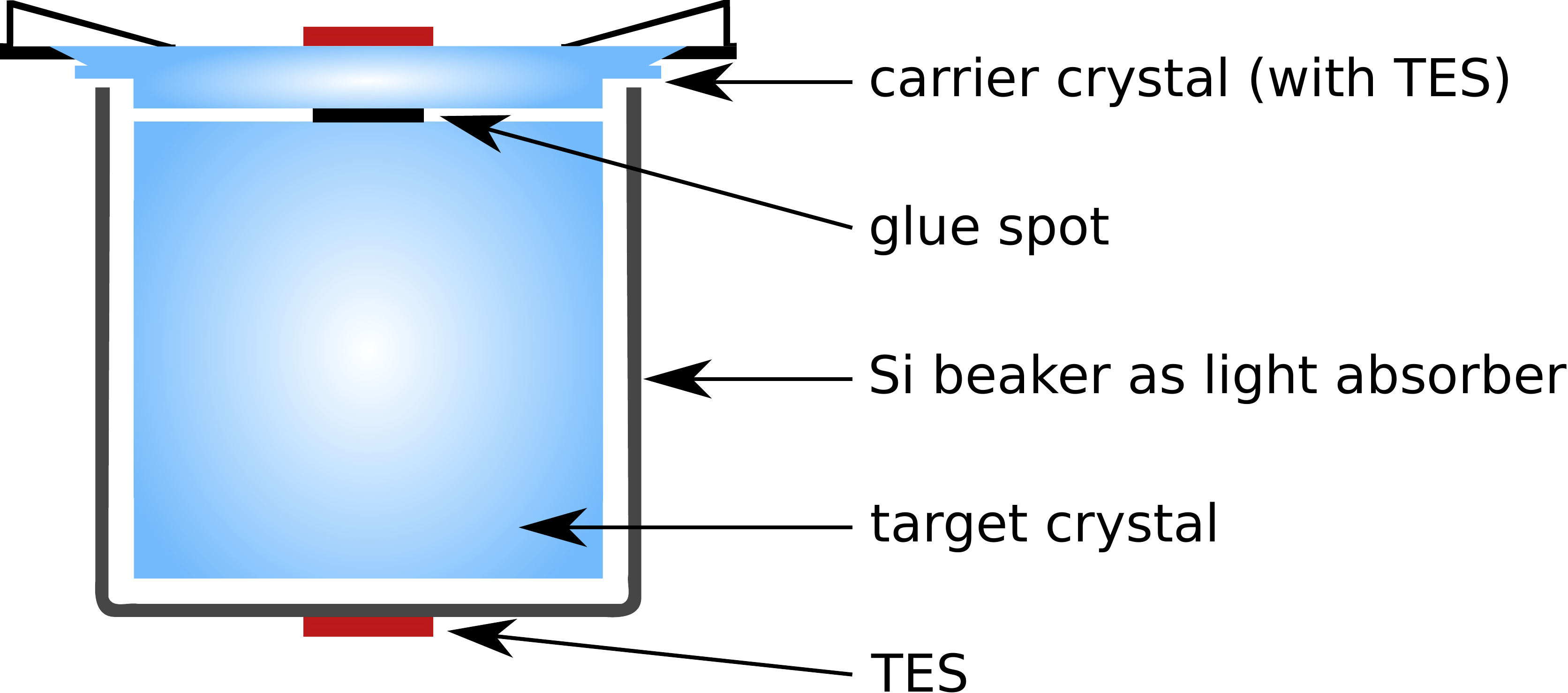}
\caption{Schematic view of the complete beaker module.
The scintillating target crystal is surrounded by a beaker-shaped silicon light absorber (area $\approx\unit[60]{cm^2}$).
The carrier crystal, on which the sensitive thermometer (TES) is located on, is shaped in such a way that any line-of-sight between target crystal, detector housing and holding structure is prevented.
Combined with the light detector a true $4\pi$ veto is established which allows to identify external and surface backgrounds.}
\label{figure_beaker_schematic}
\end{figure}

The light detector used for this detector concept has the shape of a beaker (``beaker design'').
Machined from mono-crystalline silicon, the beaker-shaped light absorber is \unit[40]{mm} in diameter and height and has a wall thickness of \unit[0.4]{mm}.
The total mass is $\approx\unit[5.9]{g}$.
The surface area the beaker offers for light absorption is $\approx\unit[60]{cm^2}$.
Temperature changes caused by energy depositions in the light absorber are measured with a tungsten TES directly evaporated onto its surface.
\begin{figure}
 \centering
 \includegraphics[width=0.750\linewidth]{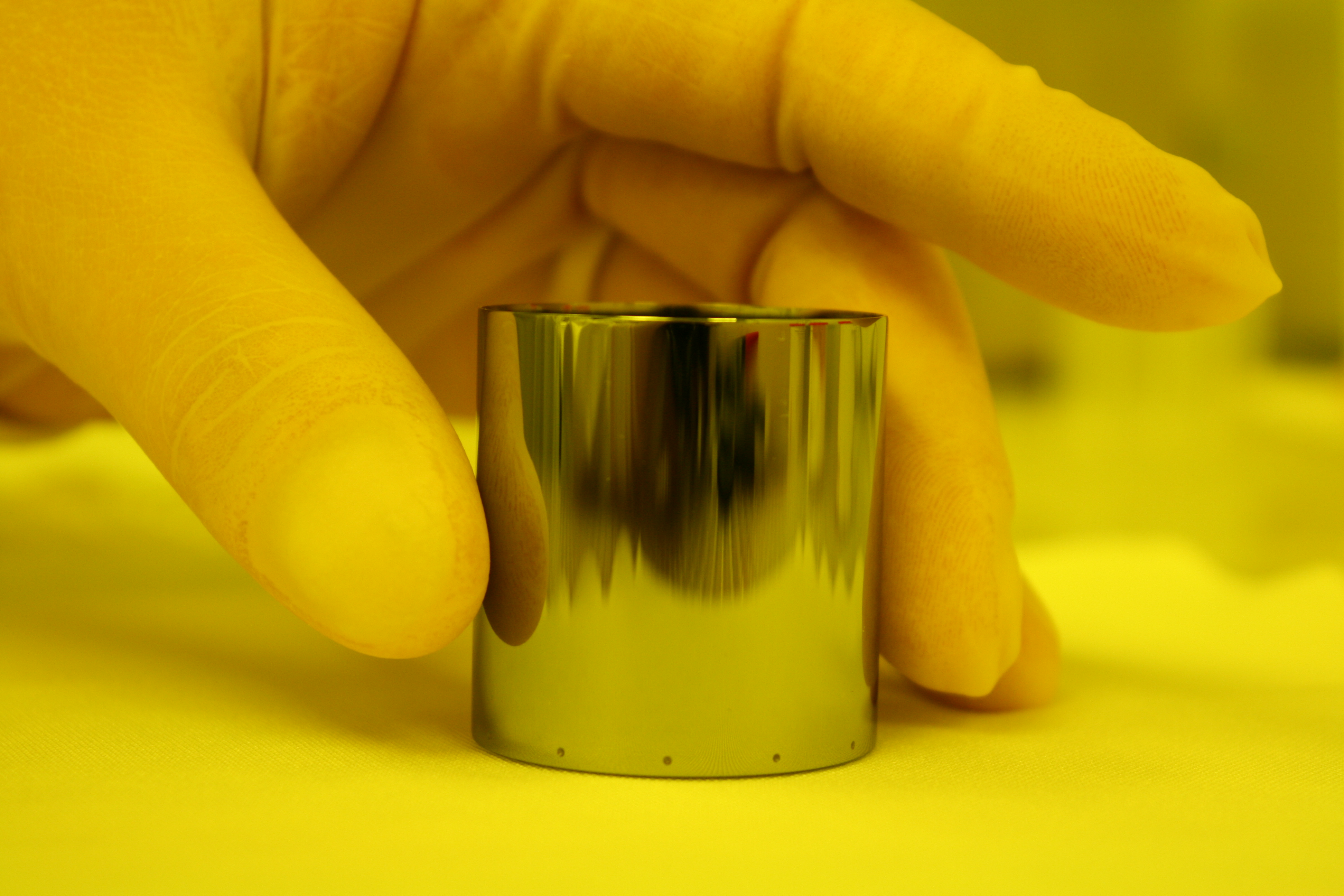}
 \caption{Photo of the beaker shaped light absorber with a diameter and height of \unit[40]{mm}. 
 The mass of the beaker is $\approx\unit[5.9]{g}$.}
\end{figure}
A composite design \cite{0912.0170v1}, i.e. the usage of a glued carrier crystal on which the tungsten TES for the phonon channel is located, is adopted from other CRESST-II detector designs in an optimized fashion. 
The carrier crystal is shaped in such a way that, in combination with the beaker, any direct line-of-sight between the target crystal and non-active surfaces is prevented ($4\pi$-veto). 
The connection between target and carrier crystals is established by a glue spot (epoxy resin) while the carrier crystal is mechanically fixed to the detector housing using bronze clamps.
Additionally, the detector housing is fully lined with scintillating and reflective VM2002\textcopyright~foil, which enhances the light collection adding a reflective surface behind the transparent carrier crystal.
The detector setup is sketched in Figure \ref{figure_beaker_schematic} (without surrounding foil).

\section{Performance of the Detector Concept}
The performance of the new detector concept is studied using data recorded by two such detector modules during the measurement campaign of CRESST-II (Phase 2) which lasted two years allowing for a high-statistics analysis.

\subsection{Light Detector as Cryogenic Detector}
\label{subsec_light_detector}

Since the dimensions of the light absorber are considerably increased compared to standard large, wafer-like light detectors, the phonons produced by an energy deposition have to undergo many surface reflections before being absorbed in the thermometer, leading to a partial thermalization in the absorber and, therefore, to a reduction of signal.
Thus, performance gains realized by the higher light collection compete with the performance reductions of the light channel as cryogenic detector.

To calibrate the light detectors an uncollimated $^{55}$Fe X-ray source illuminates the mantle of each beaker. 
The 1$\sigma$ resolution achieved for the 5.89~keV ($^{55}$Mn, K$_{\alpha}$) peak is 69.2$\pm$\unit[1.2]{eV}(0.9\%) for beaker~1 and 51.5$\pm$\unit[1.4]{eV}(1.1\%) for beaker~2.
The baseline resolution, i.e. the resolution at zero energy, is evaluated on randomly sampled empty baselines (e.g. noise samples). 
The method, thoroughly described in \cite{Lise_Results}, yields a 1$\sigma$-resolution of the baseline of $\sigma_{b,l}=5.81\pm \unit[0.05]{eV}$ for beaker 1 and $\sigma_{b,l}=7.52\pm\unit[0.06]{eV}$ for beaker 2.
The difference between the baseline resolution and the resolution obtained for the x-ray lines of the $^{55}$Fe source can be explained by uncertainties introduced by the phonon propagation in the absorber which increase with energy.

Compared to conventional, wafer-based CRESST-II light detectors, the baseline resolution $\sigma_{b,l}$ as well as the energy resolution at \unit[5.89]{keV} are on the same level.
Thus, the increased dimensions of the beaker are not reflected in the performance as cryogenic particle detector.

\subsection{Energy Detected as Scintillation Light}
\label{subsec_abs_LY}

In this context, the energy detected as scintillation light is given as the ratio of energy received in the light channel per energy measured in the absorber crystal. 
The signals of direct X-ray hits of well-known energy ($^{55}$Fe source) are compared to scintillation signals induced by \unit[122]{keV} $\gamma$-rays (external $^{57}$Co source) to set the energy scale in each absorber.
This allows to quantify the fraction of energy measured as scintillation light to the total amount of energy deposited in the absorber crystal.
Since the scintillation process is significantly slower than a direct energy deposition, pulse shape differences are observed for the different event classes. 
To take into account differences between them, the integral ratio of the respective event classes is considered when the pulse heights are compared.
CaWO$_4$ crystals translate between 7.4 and 9.2\% of deposited energy to scintillation light \cite{2016_Kiefer_In-situ}\cite{Tretjak}.
In the beaker detector design, the share of energy detected as scintillation light is $4.53\pm0.13\%$ for module 1 and $5.17\pm0.15\%$ for module 2.
Thus, the beaker design exceeds the average value of 1.95$\pm$0.3\% achieved in conventional design by factor of $>2.5$ \cite{2016_Kiefer_In-situ}.
The main reason of this improvement is the enhanced light collection efficiency of the beaker design as indicated by a dedicated simulation tracking the path of created photons in a given detector geometry which yields a value of \unit[79]{\%} for beaker module design.
Replacing the reflective foil as in a conventional design by light absorbing surfaces, the average distance photons travel inside the detector housing before being detected in the light detector is reduced by a factor of $\approx$5-10.
As a consequence, losses because of (re-)absorption in the crystal (-(75-85)\%) or by the foil (-70\%) are significantly reduced.

A large uncertainty of the simulation's output is the strong influence of the crystal quality which varies significantly between individual target crystals.
Though, the availability of the measured light collection efficiency for conventional CRESST detector designs (22.4$\pm$3.2\%\cite{2016_Kiefer_In-situ}) allows to verify the correctness of the simulation output.

\subsection{Particle Identification}
\label{subsec_particle_identification}

In the following, only module 2 is considered.
The analysis of module 1 is impeded by electronic noise of the SQUID readout amplifier, which reduces the performance of the phonon channel.

\begin{figure}
\centering
\includegraphics[width=1.0\linewidth]{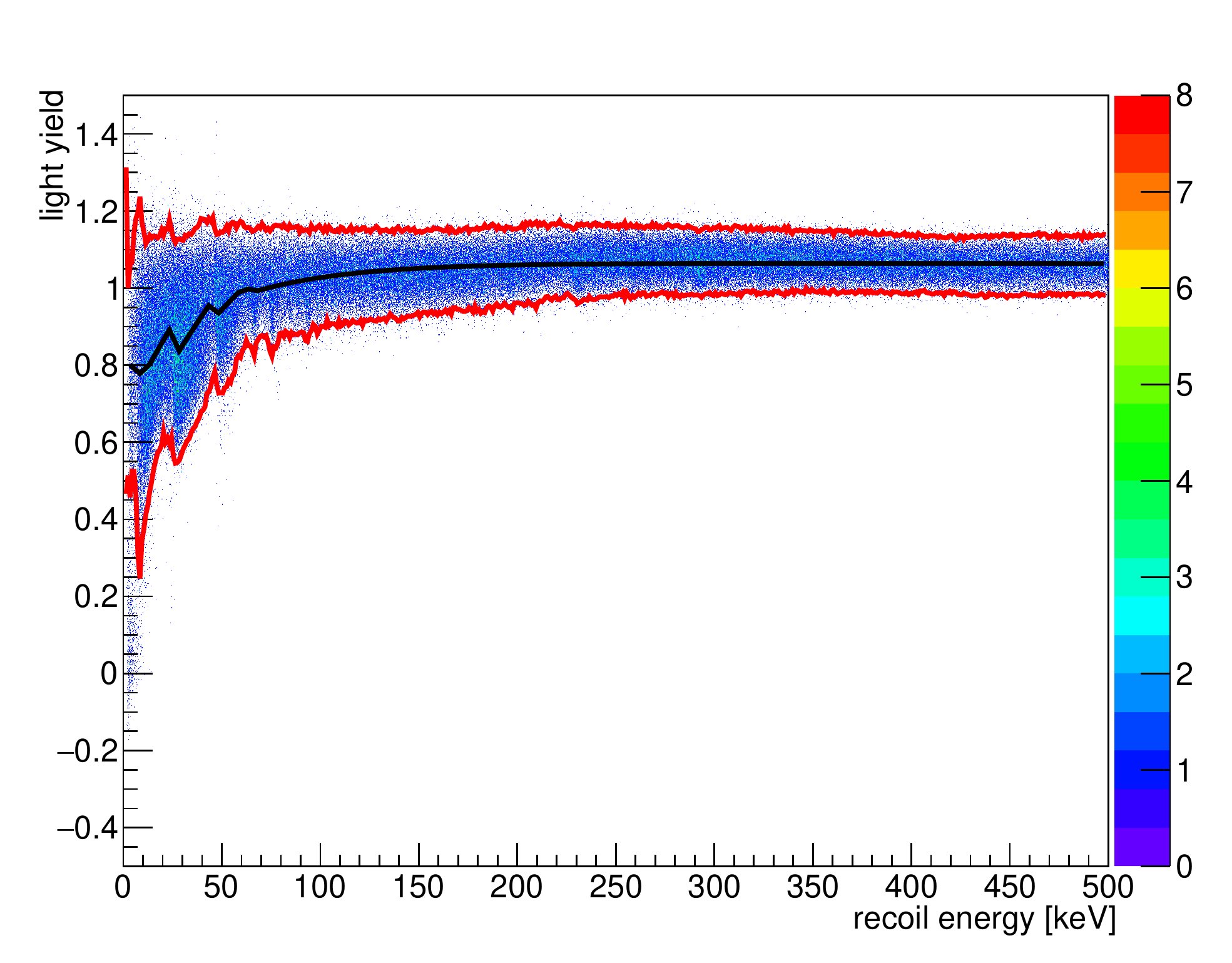}
\caption{Light yield vs. recoil energy  from  0 to \unit[500]{keV} after the application of all quality cuts. 
The $\beta/\gamma$-band is populated with internal and external backgrounds.
The black solid line shows the center value of the band while the red lines indicate the interval in which 90\% of the events are located.
The effect of the light yield band bending down towards smaller recoil energies is known as non-proportionality effect.}
\label{picture_ly_vs_energy_Vk28}
\end{figure}

As common for experiments using the phonon-light technique, the results are presented in light yield vs. energy plane. 
The light yield is defined as the ratio of energy detected in the light channel to energy detected in the phonon channel and is normalized to one for $\gamma$-events (at \unit[122]{keV}).
Nuclear recoils are quenched compared to $\beta/\gamma$ events according to their quenching factors: O=0.112, Ca=0.059, W=0.017\cite{quenching_factors}.
Figure \ref{picture_ly_vs_energy_Vk28} shows the complete data set of module 1 in the energy range \unit[0-500]{keV} after applying all quality cuts.
Additionally, the center of the $\beta/\gamma$ light yield band ($LY_c$) is indicated (black solid line) as well as the interval in which 90\% of these events are located (red solid lines).

Module 2 achieved a threshold of 1.6$\pm$0.3~keV for the phonon channel.
The value is determined by probing the region of the threshold with electrically injected heater pulses as described in \cite{Lise_Results}. 
The uncertainty is given by the baseline resolution $\sigma_{b,p}$ of the phonon channel.

The $\beta/\gamma$-band is highly populated over the entire energy range by internal and external backgrounds \cite{Radiopurtiy_Muenster}\cite{TUM40_background}.
The observed effect that the $\beta/\gamma$-band bends downwards at smaller recoil energies is known as non-proportionality effect \cite{quenching_factors}.
By fitting a single Gaussian to a given energy interval, the center and the width of the light yield distribution is determined allowing the parametrization of these quantities as function of the recoil energy $E_r$ (binned likelihood analysis; more information on the parametrization can be found in \cite{quenching_factors}).
This allows to disentangle crystal properties and light detector performance and to compare different detector modules.
Equation \ref{equation_sigma_l} describes the resolution of the light detector $\sigma_L$ as a function of the deposited energy $E_r$.
\begin{equation}\label{equation_sigma_l}
 \sigma_L(E_r)=\sqrt{S_0+S_1\cdot LY_{c}\cdot E_r+S_2\cdot (LY_{c}\cdot E_r)^2}
\end{equation}
$\sigma_L$ consist of three terms which dominate the total resolution in different energy ranges.
The parameter $S_0$ accounts for the baseline noise of both readout channels ($\sigma_{b,p},\sigma_{b,l}$) and dominates the total resolution for energy depositions $E_r<\unit[1]{keV}$.
Since the baseline noise can be derived by the independent analysis of empty baseline samples this parameter is fixed during the fit.
Additionally, the baseline resolution of the light detector $\sigma_{b,l}$ is translated in the energy scale of electron interactions taking place in the main absorber\footnote{A quantity $X$ which is translated in to the electron equivalent scale is denoted with $X_{ee}$, e.g. $\sigma_{b,l,ee}=\sigma_{b,l}/\text{energy detected as light})$}.
Thus, the parameter $S_0$ can be expressed as  $S_0=\sigma_{b,l,ee}^2+\sigma_{b,p,ee}^2$.
$S_1$ accounts for photon statistics (i.e. scales proportional with the number of detected photons) and dominates $\sigma_L$ for recoil energies in the range (1-750)~keV.
For higher energies the width of the light yield band is dominated by the term $S_2$, which accounts for possible position dependencies in the amount if detected light.
This quantity is usually small  ($mathcal{O}(10^{-3})$) and impacts the fit only at very high energies.
The results of the parametric band fit are summarized in Table \ref{table_results_band_fit}.
\begin{table}[h!]
\centering
\begin{tabular}{|c|c|} \hline
 & beaker module \\ \hline \hline
$S_0$ & $(0.103\pm0.007)$ keV$_{ee}^2$ (f) \\ \hline 
$S_1$ & $(0.318\pm 0.013)$ keV$_{ee}$ \\ \hline
$S_2$ & $<10^{-6}$ \\ \hline \hline
  & CRESST-II (avg.) \cite{schmaler_phd}\cite{zoller2016artificial}\\ \hline
 $S_0$ & $(0.635\pm0.21)$ keV$_{ee}^2$ (f) \\ \hline 
 $S_1$ & $(0.583\pm 0.085)$ keV$_{ee}$ \\ \hline
 $S_2$ & $2.98\cdot 10^{-3}\pm 2.345\cdot 10^{-4}$ \\ \hline 
\end{tabular}
\caption{Fit result after applying the described model to the data.
The parameter $S_0$ is fixed during the fit routine (indicated by (f)).
The uncertainties given are fit uncertainties (upper table) or statistical uncertainties (lower table).
}
\label{table_results_band_fit}
\end{table}

All fit parameters of the beaker module show an improvement with respect to the average values obtained for CRESST-II detector modules.
The improvements in $S_0$ and $S_1$ are directly correlated to the increased light collection efficiency. 
This can be confirmed by the observation that $S_0$ and $S_1$, if corrected for the improvements correlated to the higher light collection efficiency, are very close to the average values obtained for CRESST-II light detectors.
A reduction of $S_0$ by $\approx 60\%$ is expected based on the higher light collection efficiency if one assumes the same cryogenic performance of the light detector.
Furthermore, $S_1$ scales inversely with the amount of detected light.
Thus, an improvement of a factor 2.5 is expected for $S_1$.
A different effect accounts for the improvement of $S_2$.
From the simulation used to estimate the light collection efficiency of the beaker module (see section \ref{subsec_abs_LY}) we derive a 5 times shorter average distance photons travel before reaching the light detector.
As a consequence, the position of the energy deposition impacts the amount of detected photons less since losses during the photon propagation are less probable (-(75-85)\%).
This is reflected in the very small value of $S_2$.

The improved energy resolution of the light channel is reflected the good discrimination power of the beaker module.
The expected leakage of $\beta/\gamma$-background into the acceptance region of dark matter searches (\unit[0-40]{keV}; LY$<0.112$ (i.e. center of the oxygen recoil band)) is reduced to 51$\pm$4\% compared to an average conventional CRESST-II detector module.
All improvements can be attributed to the geometry while a degradation of the light detector performance due to the enlargement of the light absorber is not observed.

\subsection{4$\pi$-veto System}
\label{subsec_veto_system}

The main motivation of the beaker design is the unambiguous identification of surface events and external backgrounds.
For CRESST-II two main classes of external backgrounds are identified which limit the sensitivity of the experiment: Surface related $\alpha$-background and excess light events.
Additionally, the identification of carrier related events is mandatory since this part of the detector is in direct contact with non-scintillating surfaces.
In the following, all different event classes are studied.

\subsubsection{Carrier Events}
The identification of \textit{carrier events} is achieved by exploiting the pulse shape differences which result from the difference in absorber volume (for more information see \cite{Proebst95}).

\begin{figure}
\centering
\includegraphics[width=1.0\linewidth]{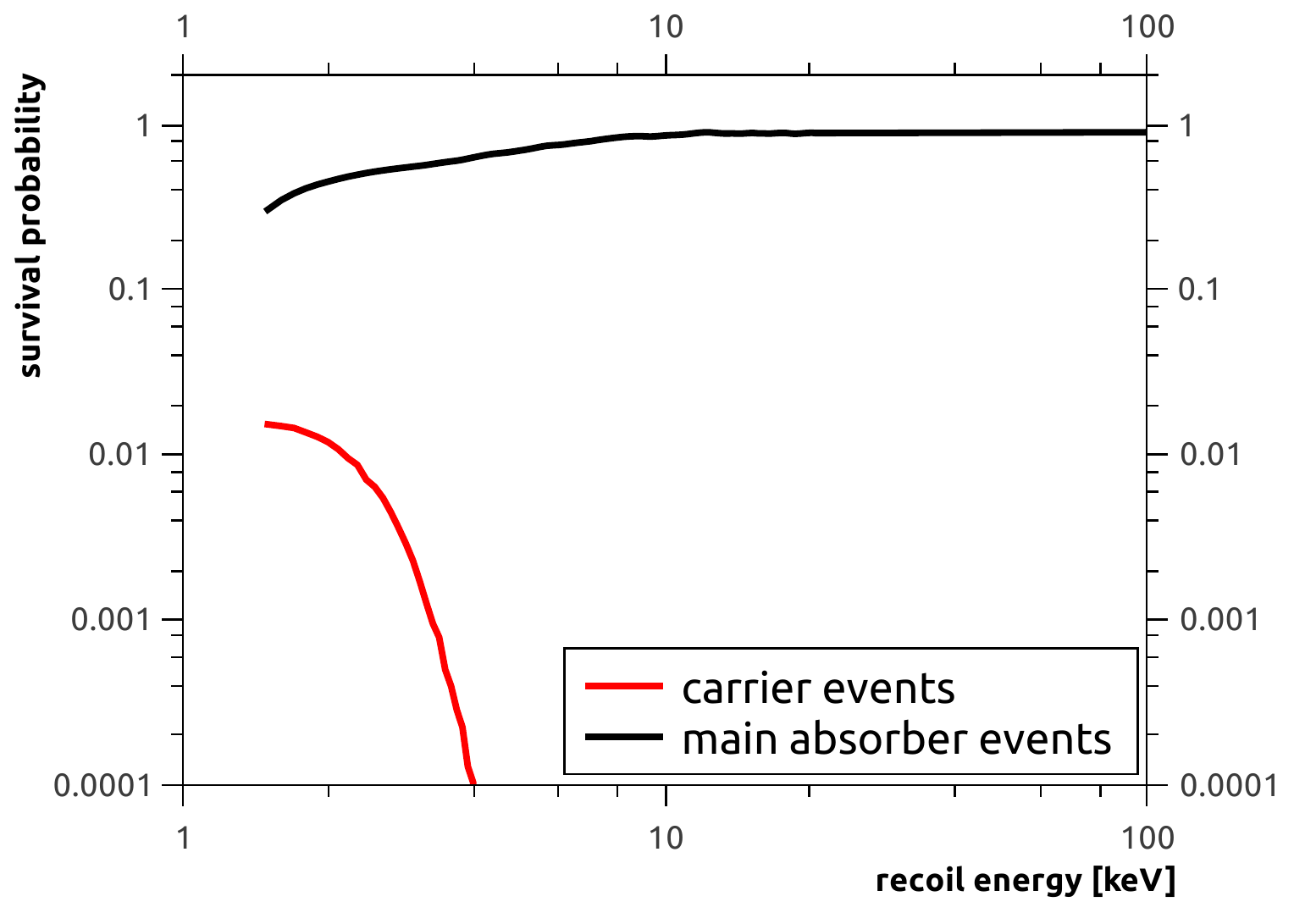}
\caption{Survival probability for simulated main absorber and carrier-like events after all cuts were applied.
While the survival probability of main absorber events is only insignificantly reduced below \unit[10]{keV}, the carrier discrimination power is $>10^8$ for energies above \unit[5.4]{keV}.
For lower energies the survival probability of carrier events grows and reaches its weakest discrimination power at threshold ($1.52\cdot 10^{-2}$).
For the given data set less than 1 carrier event is expected to survive the pulse shape discrimination.}
\label{pic_survival_prob}
\end{figure}

To determine the cut efficiency for a certain event class, the survival probability is calculated by applying the cuts on simulated events (method described in \cite{Lise_Results}). 
The result of this procedure is depicted in Fig. \ref{pic_survival_prob} for main absorber and carrier-like events, giving the survival probability of these two event classes.
The simulated data show that for energies above \unit[5.4]{keV} the survival probability for carrier-like events is $<10^{-8}$.
Below \unit[5.4]{keV} the survival probability grows and reaches its maximum at threshold with $1.52\cdot 10^{-2}$. 
Considering the full data set less than one carrier event is expected to survive the pulse shape cuts in the region of interest for dark matter search.

\begin{figure}
 \centering
 \includegraphics[width=0.9\linewidth]{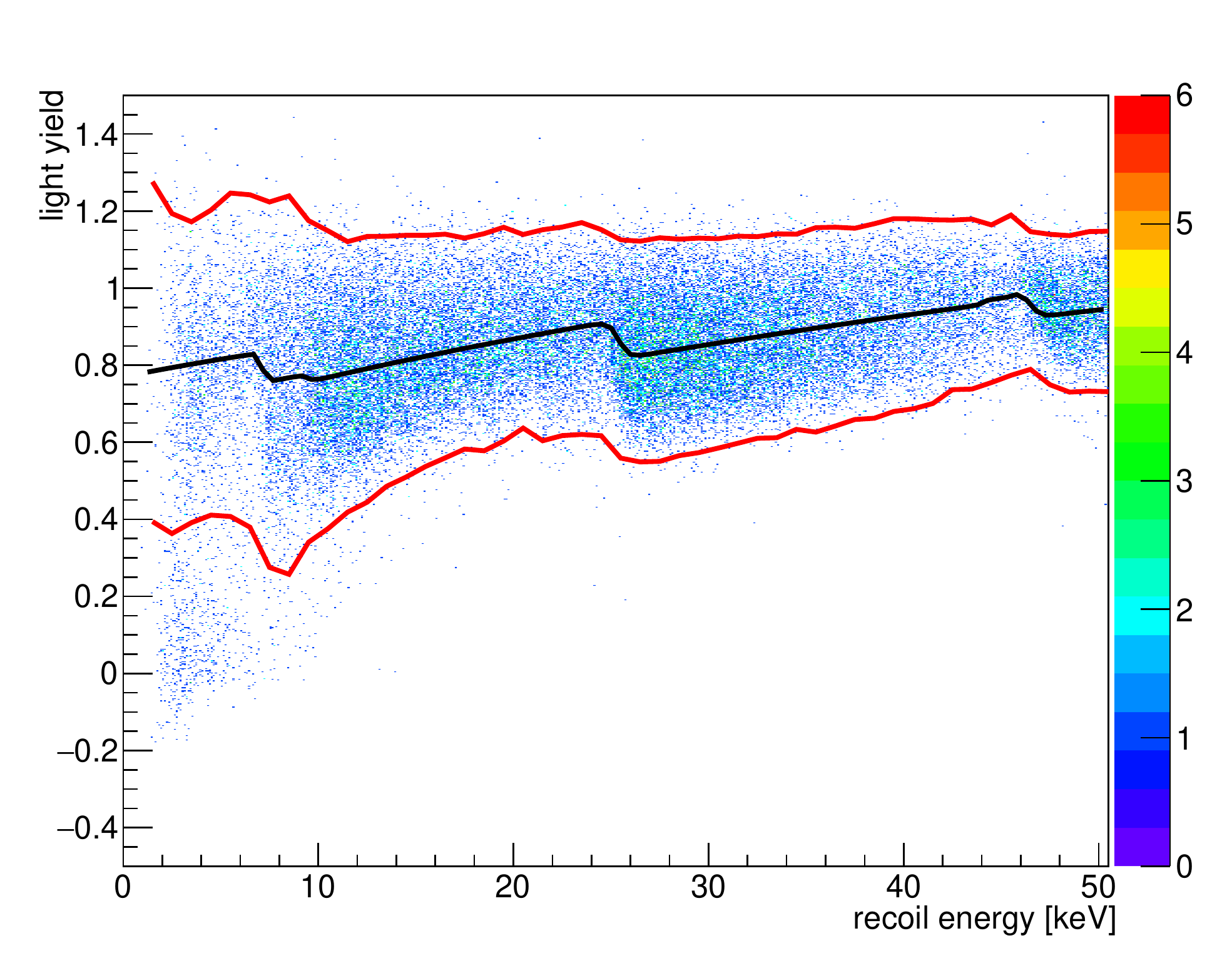}
 \caption{Light yield vs. recoil energy scatter plot between 0 and 50~keV after the application of all quality cuts.
 Additional to the internal backgrounds expected to arise in the $\beta/\gamma$-band, an event population with reduced light yield is visible.
 Starting from $\approx\unit[10]{keV}$, this population arises with an average light yield of 0.
 A correlation to the carrier system is likely since stress-relaxation in the glue spot can cause this event population.
 }
 \label{VK28_scatter_50_keV}
\end{figure}

\subsubsection{No-light Events}
Below $E_r<\unit[10]{keV}$ the data set shows an additional event population which features a mean light yield of zero (see Figure \ref{VK28_scatter_50_keV}).
Misidentified carrier events as explanation for this event class is not plausible since pulse shape analysis of the phonon channel vetoes carrier events effectively in this energy region.
Indeed, the pulse shape of such \textit{light-yield-zero events} is identical to main absorber events. 
However, a correlation with the carrier system cannot be excluded. 
Stress relaxation in the intermediate glue layer may be a source of light-yield-zero events, mimicking nuclear recoils in the region of interest.
This explanation is additionally motivated by the large glue spot which is necessary to ensure a proper mechanical connection between the carrier and the main absorber guarantees an efficient phonon transfer. 

\subsubsection{Excess-light events}
External $\beta/\gamma$-backgrounds (which originate from detector housing and cryostat walls) can be identified using two different mechanisms.
First, by an additional energy contribution in the light channel and, secondly, by using pulse shape analysis. 
The additional energy deposition in the light detector is caused by $\beta/\gamma$-particles traversing the beaker. 
They can be either fully stopped in the light detector absorber or deposit there a fraction of their energy before being stopped in the phonon detector.
In this last case the resulting event class exceeds the light yield of an internal $\beta/\gamma$-event at the same recoil energy $E_r$ and is therefore called  \textit{excess light event.}
Additionally, as part of the energy is deposited directly in the light channel, the pulse shape of this event class differs from pure scintillation events (see section \ref{subsec_abs_LY}) allowing pulse shape analysis to identify them.

\subsubsection{Surface events}
The unambiguous identification of surface related $\alpha$-back-\\ground is guaranteed since the detection of the full energy (i.e. Q value) is assured.
For CRESST-II detectors the most prominent surface background were $^{210}$Po $\alpha$-decays (Q value: \unit[5307]{keV} ($^{206}$Pb nucleus (\unit[103]{keV}) + $\alpha$-particle (\unit[5204]{keV} )))\cite{Firestone}, induced by the progenies of $^{222}$Rn surface contaminations during detector production and installation.
Depending on how the energy is distributed to the different detector parts (light detector, phonon detector or carrier), different signal signatures are observed.
In case the $\alpha$-particle deposits its energy in the main absorber while the $^{206}$Pb nucleus deposits its energy in the beaker, the event is tagged by the additional energy the $^{206}$Pb nucleus deposits in the light channel. Since the total energy deposition in the light channel is significantly enhanced by the absorbed $^{206}$Pb nucleus, the normally achieved light yield of $\beta/\gamma$ events is exceed by $\mathcal{O}(10^2)$ which allows an unambiguous identification.

If the $^{206}$Pb nucleus is detected in the main absorber while the $\alpha$-particle escapes, the observed detector response is different.
The energy observed in the phonon channel is small and is combined with the quenched light yield of a nuclear recoil \cite{quenching_factors}.
The identification of this event as surface event relies on the detection of the escaping $\alpha$-particle in the beaker which adds to the scintillation signal.
Although surface $\alpha$-events are classified unambiguously as background, a precise quantification of the rejection capability is difficult.
The energy depositions in the light detectors are well above the dynamic range of their TES, therefore pulses are saturated which prohibits pulse shape analysis to disentangle different event classes.
Still, a rough estimate using the energy signature measured by the phonon channel can be given for surface-related $^{210}$Po $\alpha$-decays.
For both scenarios explained above, the recoil energy in the main absorber is unambiguous, which allows to isolate surface events by their extreme light yield.
For the full data set we can assign $\mathcal{O}$(1000) events to this background process while not a single untagged event is observed.
The distinct identification of surface events with degraded nature \footnote{Degraded refers to either the $\alpha$-particle or the recoiling nucleus not depositing their full energy in the main absorber\cite{CRESST12}\cite{sputter}.} is impeded by the presence of other backgrounds (e.g. \textit{excess light events}).
In region of interest of dark matter searches, though, not a single surface-related event is observed after the application of cuts while in conventional detector modules $\mathcal{O}(10)$ of such events remain\cite{CRESST12}.

\section{Conclusion}

The beaker design is a promising detector concept which exploits a large scale beaker-shaped light detector in order to achieve a better particle discrimination and to establish $4\pi$-veto.
Both design goals are confirmed by a proof-of principle measurement performed during CRESST-II (Phase~2).
The large active area increases the energy detected as light by a factor of 2.5 compared to other CRESST-II detector designs. 
From an average value of $\approx\unit[2]{\%}$, the absolute light yield is raised to an value of $\approx\unit[5]{\%}$ achieved in the beaker design.
This improvement is mainly related to the improved light collection inside the detector module while the performance of the light detector as cryogenic device does not suffer from enlargement of the light absorber.
The achieved baseline resolutions of $5.81\pm \unit[0.05]{eV}$ and $7.52\pm\unit[0.06]{eV}$ show no degradation in performance compared to smaller CRESST-II light detectors.

Because of the larger light collection, background identification is enhanced for the region of interest of dark matter searches.
The parameters describing the light yield bands are improved as expected for a better light collection efficiency.
Furthermore, the influence of position dependencies is reduced, since photons have to travel shorter distances before being absorbed in the light detector.

The $4\pi$-veto system proved its capability to identify known background sources. 
External backgrounds and surface $\alpha$-events are discriminated by their unique signatures in the respective detector parts with high efficiency. 
The identification of carrier events, crucial for establishing a true $4\pi$-veto, is achieved down to detection threshold.

Currently (in CRESST-III (Phase 1)), an improved version of the beaker concept is tested.
Aiming at the investigation of the event population with light yield zero, this detector concept is equipped with a second phonon read-out channel.
Assuming that the origin of this light yield zero contribution is the non-scintillating glue, the ratio of both channels is expected to allow the disentanglement of main absorber events and events with different origin.

Finally , we want to emphasized that the beaker concept can be applied to any rare event search using cryogenic techniques. 
Thanks to its outstanding vetoing capabilities the beaker concept offers a "background-free" detector environment for all types of target crystals (scintillating or non-scintillating).

\section{Acknowledgements}
We are grateful to LNGS for their generous support of the CRESST experiment.
This work was supported by the DFG cluster of excellence: Origin and Structure of the Universe, by the Helmholtz Alliance for Astroparticle Physics, and by the BMBF: Project 05A11WOC EURECAXENON.

\bibliographystyle{unsrt}
\bibliography{bibliography.bib}

\end{document}